\documentclass[sigconf]{acmart}

\AtBeginDocument{%
  \providecommand\BibTeX{{%
    \normalfont B\kern-0.5em{\scshape i\kern-0.25em b}\kern-0.8em\TeX}}}
    
\begin{document}

\setcopyright{acmcopyright}
\copyrightyear{2020}
\acmYear{2020}

\acmConference[FDG'20]{FDG'20: Foundations of Digital Games}{September 15--18, 2020}{Malta}
\acmBooktitle{FDG'20: Foundations of Digital Games,
 September 15--18, 2020, Malta}

%
\title{Towards Friendly Mixed Initiative Procedural Content Generation: Three Pillars of Industry}
%

\author{Gorm Lai}
\affiliation{%
 \institution{Goldsmiths, University of London}
 \city{New Cross, London}
 \country{United Kingdom}}
\email{glai001@gold.ac.uk}
 
\author{William Latham}
\affiliation{%
 \institution{Goldsmiths, University of London}
 \city{New Cross, London}
 \country{United Kingdom}}
\email{w.latham@gold.ac.uk}

\author{Frederic Fol Leymarie}
\affiliation{%
 \institution{Goldsmiths, University of London}
 \city{New Cross, London}
 \country{United Kingdom}}
\email{ffl@gold.ac.uk}

%
\renewcommand{\shortauthors}{Lai et al.}

%
\begin{abstract}

  While the games industry is moving towards procedural content generation (PCG) with tools available under popular platforms such as Unreal, Unity or Houdini, and video game titles like No Man's Sky and Horizon Zero Dawn taking advantage of PCG, the gap between academia and industry is as wide as it has ever been, in terms of communication and sharing methods. One of the authors, has worked on both sides of this gap and in an effort to shorten it and increase the synergy between the two sectors, has identified three design pillars for PCG using mixed-initiative interfaces. The three pillars are \emph{Respect Designer Control}, \emph{Respect the Creative Process} and \emph{Respect Existing Work Processes}. \emph{Respecting designer control} is about creating a tool that gives enough control to bring out the designer's vision. \emph{Respecting the creative process} concerns itself with having a feedback loop that is short  enough, that the creative process is not disturbed. \emph{Respecting existing work processes} means that a PCG tool should plug in easily to existing asset pipelines. As academics and communicators, it is surprising that publications often do not describe ways for developers to use our work or lack considerations for how a piece of work might fit into existing content pipelines.
\end{abstract}

%
%



%
\maketitle

\section{Introduction}
\label{sec:introduction}

In recent years, both academia and the games industry have had an increased focus on PCG. However, except for isolated attempts such as Natural Motion~\cite{company:NaturalMotion}, we need more effort in bringing academic research to use as inspirational or developmental material for the games industry. This is an opinion also expressed in Shaker et al.~\cite{Doull2014} in their interview with Andrew Doull, as he states:``There's a lot of interesting stuff happening on the academic side - getting this to percolate over to game development is going to be the real challenge.'' In game AI Togelius~\cite{Togelius2014}, and from the wider tech-industry Kaczmarczyk~\cite{Kaczmarczyk2015}, have both lamented the lack of collaboration and understanding between academia and the industry. With 16 years working in the games industry, from indie-sized to AAA productions, one of the authors of this paper has the same experience.

The main contributions of this work are to distill such issues into three design pillars for creating tools for mixed-initiative procedural content generation (MI-PCG): \emph{Respect Designer Control}, \emph{Respect the Creative Process} and \emph{Respect Existing Workflow}.

\emph{Respect designer control} focuses on empowering the designer to be able to get their vision out. It asks the question, ``does the algorithm provide enough control for the designer to express their vision?''  \emph{Respect the creative process} concerns itself with providing a short iterative loop that provides enough feedback to the user that it does not break their creative process. The last pillar, \emph{respect existing work processes} focuses on the ease of embedding PCG tools into an organisation with an existing workflow that already combines a number of other tools. It is important to figure out exactly where a new tool fits into the workflow, who provides data for it, where the generated content goes next, and what and who is affected when the content is iterated upon. After introducing the pillars by referencing existing literature, we argue in case studies in sections~\ref{CaseStudy1} and~\ref{CaseStudy2} that these pillars are in fact useful to industry.

It is important to note we are not arguing other approaches to MI-PCG are not valid. The position as it is presented here, is that if you want to make a MI-PCG tool which considers an industry audience, then we put forward arguments that recommend you consider our three pillars. Research with a different focus exist outside this position and is not contrary to it. 

\section{Background}
\label{sec:previous}

In their book~\cite{Norman1986} Norman et al. introduce the two concepts, \emph{the gulf of evaluation} and \emph{the gulf of execution}. The gulf of evaluation is how easy it is for a user to perceive representations and feedback, and the gulf of execution is how easy it is to accomplish a given task. The gulf of execution is an important concept for respecting designer control. Shortening the gulf of evaluation and the gulf of execution are important parts of respecting the creative process as both are part of the iterative creative process.

Moving on from general user experience to mixed-initiative (MI) interfaces, in one of the earliest publications on the subject Horvitz~\cite{Horvitz1999} lists twelve principles for MI interfaces with direct manipulation. The first of these deals with the question, ``is the tool actually adding value?'' A MI interface need to add tangible advantages over a well crafted user interface with no agent.

Building on that, Liapis et al.~\cite{Liapis2016} lists six requirements and questions for designing MI-PCG tools. The one question that is most relevant to our work is ``How can the method for control over content be balanced?'', has a clear answer in the context of this paper, as we will see in the case study in section~\ref{CaseStudy1} and in the analysis in section~\ref{sec_RespectDesignerControl}. The follow up question, ``How to resolve conflicts that arise due to the human stating conflicting desires?'', leads us invariably to the question of how can we give maximum control to the user without resorting to manual labour, or similarly, how can we create the most empathetic agent possible? 

Regarding \emph{respect existing work processes}, the authors have been able to find little literature on the analysis of asset pipelines or on MI interfaces that describe how such tools integrate with the workflow of the average game developer. In lieu of not being able to find many references we have made a case study (section~\ref{CaseStudy2}), that will argue how important it is for a content creation (CC) tool to fit into existing work processes and asset pipelines.  


\section{The Three Pillars}
\subsection{Respect Designer Control}
\label{sec_RespectDesignerControl}
When we invent a new tool for PCG, we must make sure that the tool gives enough control to the creator that it can bring out the designer's vision. As Lambe~\cite{Lambe2012} states ``Manual creation's strengths are PCG's weaknesses''. Manual methods give the control the designer ultimately wants (as argued in section~\ref{CaseStudy1}), but does not provide the automation and potential speed up in work process that some PCG tools can provide. Thus the impossible sounding goal is to create a tool with the speed of automation and control of manual labour. Gingold's ``magic crayons''~\cite{Gingold2003} embody this approach as they ``enable authors to obtain satisfactory results with a small amount of effort'', ``are artistically expressive'' and ``are magic because they are imbued with the power of computation.'' The MI magic crayon, is the empathetic tool that directly senses what a user is trying to do in a given context. In ~\cite{Liapis2012, Liapis2014a} Liapis et al. get closer to this MI magic crayon by trying to learn the user's preferences.

Liapis et al.~\cite{Liapis2016} describe the ``user fatigue'' that sets in, when a tool does not allow the designer to converge fast enough on the wanted result, for example, when the designer is forced to go through too many iterations, if feedback is slow, when there are too many options or the interface requires a very specific input. These boil down to providing adequate controls for the designer to do what they need to in order to achieve a given goal. We see that the width of the gulfs of execution and evaluation are important factors in causing user fatigue.

Giving control by allowing the designer to overrule the MI-PCG agent is not equal to \emph{respecting designer control}, as this takes away the co-creative human-agent relationship entirely. Instead some MI interfaces such as Tanagra~\cite{Smith2010, Smith2011} allow local editing, while the surrounding generative level geometry adapts to those changes to satisfy pre-defined constraints.

\subsection{Respect the Creative Process}
\begin{figure}[h!]
    \centering
    \includegraphics[width=0.7\linewidth]{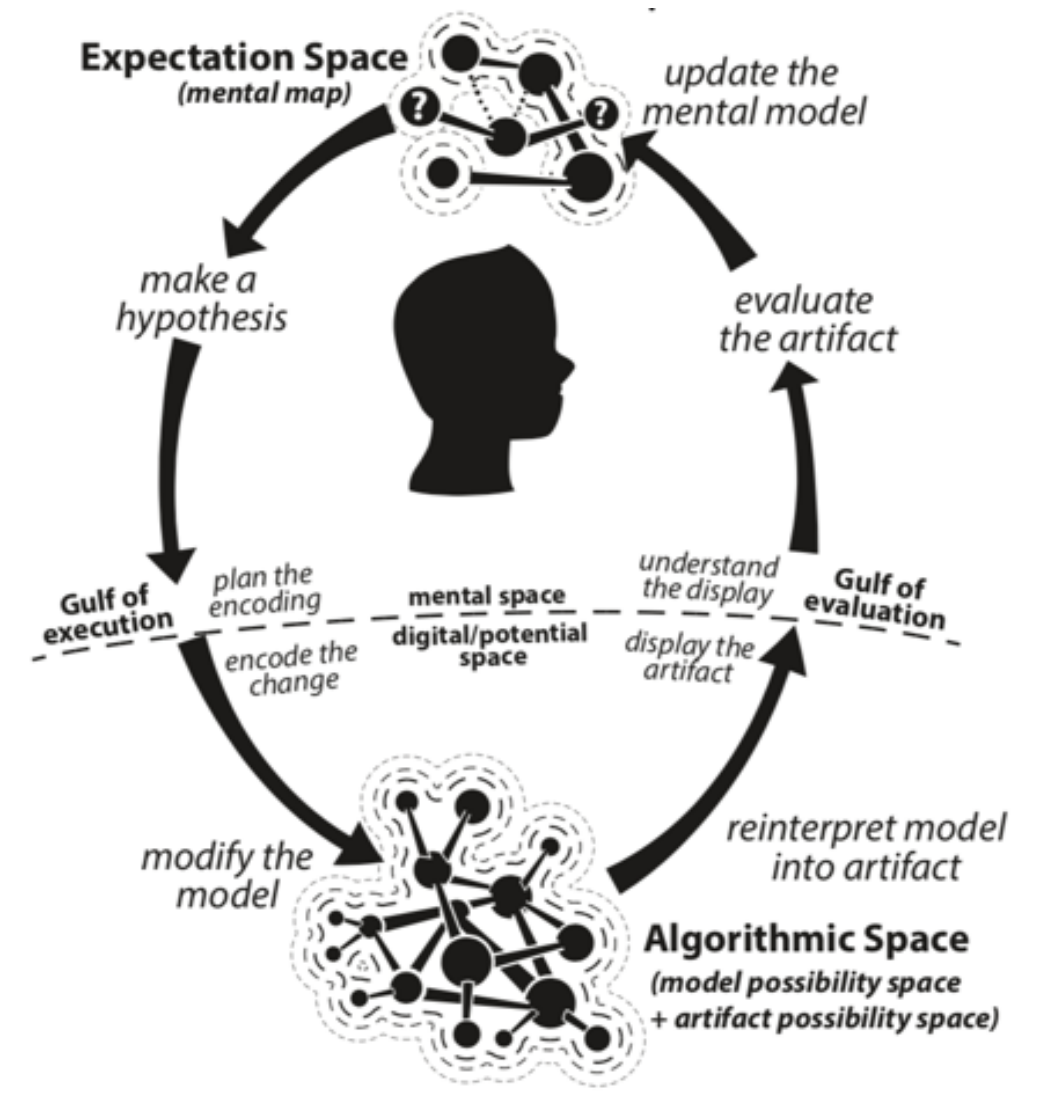}
    \caption[]{Figure 8.2 from ~\cite{Compton2019} visualising the grokloop\protect\footnotemark}
    \label{fig_compton2019_figure8dot2_grokloop}
\end{figure}

When making changes to procedurally generated content, the iteration must be so fast, that it can compete with the instant feedback of manual labour~\cite{Buttner2019_17_21, Sims1987}. The creative process is a feedback loop of trying something out, seeing the result, making changes, seeing the new result, making further changes, and so forth until the designer is satisfied. To stay focused on the task, it is important that this feedback loop is as short as possible. As shown in figure~\ref{fig_compton2019_figure8dot2_grokloop} Compton~\cite{Compton2019} distills the feedback loop into four steps: \emph{build a hypothesis}, \emph{modify the model}, \emph{evaluate the result} and \emph{update the model}. Compton defines these four steps as the ``grokloop'', and considers it as being a way for the user to interact with a generative tool and examining its possibility space. She says ``I found myself wanting a way to say \emph{the speed of learning depends on how short the loop is}''. While Compton's work is directed towards casual creators, her arguments are just as valid for game developers. \emph{Respecting the creative process} echoes Compton's call to make a grokloop that is as short as possible.

\footnotetext{Reproduced with permission of the author.}
The speed of the grokloop is not only important when interacting with an agent, but also when training an agent. This is the point where machine learning (ML) algorithms with long training times can fall short~\cite{Buttner2019_17_21}. In their breakthrough paper on interactive machine learning (IML) Fails et al.~\cite{Fails2003} introduce the principle ``fast and focused''. They focus on empowering non-technical users to quickly create classifiers for ML based image classification algorithms, and they describe the principle as upheld if the user is able to quickly create a classifier, while being focused on the classification problem itself, and not the underlying algorithms. Their variant of the grokloop is ML specific and specialises to \emph{manually classify}, \emph{generate classifier} and \emph{evaluate classifier}. Through the principle of ``fast and focused'', a successful application will be able to perform this loop at interactive frame rates and break out of it entirely within minutes. The principles of IML are in opposition to classic ML algorithms where a high level of technical skill is often required, and training data often needs to be fed to the algorithm for a long time, without an interactive interface to test when the user is happy with the result of the learning.

\subsection{Respect Existing Work Processes}

To facilitate maximum adoption of PCG tools in the industry, it is important that innovations plug directly into existing workflows. To make a tool easy and cost-effective to adopt, it should not affect the workflow of other people on the team. If two artists A and B work on the same asset at different points in the pipeline, and B is dependent on A's output, a PCG tool to optimise part of  A's workflow should output assets in a format that B can work with. It is worth remembering that while most academic research is done on an individual level or in small teams, many video games are made by teams orders of magnitude larger. For example over 3000 people worked on Red Dead Redemption 2~\cite{game:RedDeadRedemption2, RedDeadRedemption2Credits}. Even with smaller teams of tens or hundreds of people using a plethora of tools, changing workflow becomes a potentially costly and risky proposition. If a research team wants to maximise the chance their research will be adopted into a game production pipeline, they need to adopt the path of least resistance. Examples of commercial PCG tools with a focus on easy integration with existing asset pipelines can be seen in table~\ref{fig_content_pipelines}.

\section{Case Studies}

We now discuss two case studies to further argue the validity of our three pillars. The first study focuses on the improvements on the Evolutionary Dungeon Designer~\cite{Dahlskog2015} made throughout a number of works~\cite{Baldwin2017a, Baldwin2017, Alvarez2018, Alvarez2018a}. These improvements mainly relate to the first pillar, \emph{respect designer control}, though the second pillar, \emph{respect the creative process}, is also expressed. The second case study focuses on the third pillar, \emph{respect existing work processes}, by examining twelve randomly selected PCG tools, and what, if any, features they emphasise to ensure they fit into existing work processes.

\subsection{Evolutionary Dungeon Designer}
\label{CaseStudy1}
In~\cite{Dahlskog2015, Baldwin2017a, Baldwin2017, Alvarez2018, Alvarez2018a}, Dalhskog et al., Alvarez et al. and Baldwin et al. created a series of works based on the procedural generation of dungeons via a MI interface. The initial work~\cite{Dahlskog2015} analyzed 91 games published between 1975-1993 and extracted patterns, that was transformed into mechanical game design patterns to be used for procedural content. Through user surveys they made improvements documented in~\cite{Baldwin2017, Alvarez2018, Alvarez2018a}. In the first of those works~\cite{Baldwin2017}, 3 out of 5 users requested some sort of local control, so that edited parts were not modified by the evolutionary search algorithm. This supports the pillar \emph{respecting designer control}. It can be argued that the tool already upholds the first pillar, as the user is able to manually edit parts, or the user could keep generating new searches until they found something they were happy with. However, these arguments are easily dismissed. In the former case, the user would have to design the whole dungeon by hand, in which case the co-creative agent has been entirely omitted. In the latter case, in theory the user could keep using the tool to search for new layouts until they found something pleasing enough to be a starting point for manual editing. However, there is no guarantee that the algorithm would converge within reasonable time, so most likely user fatigue as described in section~\ref{sec_RespectDesignerControl} would set in. Also, 4 out of 5 users argued they would like to have a way of highlighting the design patterns. Seen through the lens of\emph{respecting the creative process}, this is a way of shortening the grokloop.

More of the suggested improvements were implemented and described by Alvarez et al. in~\cite{Alvarez2018}. Along with further improvements Alvarez et al.~\cite{Alvarez2018a} also includes a user survey, where the authors conclude:``To a certain extent, controllability is preferred than expressivity, as the users continuously try to impose their vision, which is a non-trivial task for automated systems to capture, thus, the users are more likely to sacrifice to a certain degree expressivity and exploration of the tool by gaining control over the generated content.''. This expresses the pillar, \emph{respect designer control}.

\subsection{PCG Middleware}
\label{CaseStudy2}

Table~\ref{fig_content_pipelines} show custom content pipeline support as described by the products' own advertising on the web. We have trusted this advertising and it has not been verified, as in this context, the important part is the intent that the product's description declares.

\begin{table}[t]
    \centering
    \begin{tabular}{ |l|p{47mm}| }
      \hline 
      \textbf{Product} & \textbf{Content Pipeline Support} \\
      \hline
      Arbaro\cite{application:Arbaro} & Exports to Povray, DXF and Wavefront Obj 3D model formats\\
      DreamScape\cite{application:DreamScape} & Integrates with 3D Studio Max\\ 
      Houdini Engine\cite{application:HoudiniEngine} & Integrates with common CC tools such as Maya, Cinema4D and Unity\\
      Mixamo\cite{application:Mixamo} & Exports to formats used in Unity, Unreal and Blender\\
      Speed Tree\cite{application:SpeedTree} & Integrates with Unity, Unreal and Lumberyard. Provides custom SDK\\
      Substance Designer\cite{application:SubstanceDesigner} & Integrates with Maya, Houdini,\\
      Substance Painter\cite{application:SubstancePainter} & Unity, Unreal and more\\
      Tree It\cite{application:TreeIt} & Exports to FBX, Wavefront Obj, X and more 3d model formats\\
      Vue\cite{application:Vue} & Integrates with Maya, Cinema4D, and exports to common 3d model formats\\
      World Creator\cite{application:WorldCreator} & Extensive import and export support for 3D geometry and 2D Maps\\
      World Machine\cite{application:WorldMachine} & Exports to formats compatible with common CC tools\\
      Xfrog\cite{application:XFrog} & Integrates with Maya, Cinema4D \\
      \hline
    \end{tabular} 
      \caption{Advertised asset pipeline support by PCG tools}
      \label{fig_content_pipelines}
\end{table}

Most of the larger commercial software packages list one or more sales pitches about how well their software integrates with and can be customized to a games studio's pipeline. For example on the Houdini Engine~\cite{application:HoudiniEngine} web page, it says ``Houdini Engine supports a deep integration of Houdini and its procedural workflow within the larger framework of a Studio pipeline.'' This re-assures decision makers that the tool will optimise and not disrupt work processes.

All twelve pieces of software provide some sort of integration with a studio's workflow, even if in the case of freeware like Arbaro~\cite{application:Arbaro}, it is just a simple export functionality. We conclude that providing support for integration with a studio's content pipeline is a central feature to be advertised by PCG tool-makers.

\section{Conclusion}
We have argued through referencing literature and two separate case studies, that at least three central design pillars should be taken into consideration when researchers work on MI-PCG tools which consider a games industry audience. These three pillars are \emph{respect designer control}, \emph{respect the creative process} and \emph{respect existing workflow}. While central works on MI works have focused on attributes that fall within the pillar of \emph{respect the creative process}, such as a short grokloop and visual feedback, most works disregard \emph{respect designer control} by resorting to letting the designer lock down and manually edit parts of a piece of content. This inspires us to think much can be gained by focusing on increasingly empathetic magic crayon~\cite{Gingold2003} agents.

Regarding \emph{respecting existing work processes}, we should begin to think about how PCG tools integrate with existing content pipelines. This can be done as high-level thinking about where in the content pipeline a tool fits in, who it will be used by and how the exported content will be used further down the line. These questions inform us about the possible use and important features of a tool. Knowing that artists, programmers and level designers often work on the same data at different stages in the pipeline, can affect whether we put the emphasis on art or game logic, who it affects when new edits are applied to existing data, what data the tool imports/exports and consequently what sort of data is processed and how abstractions will be modelled. On a more practical level, some tools in table~\ref{fig_content_pipelines} support data making roundtrips, so the same data can be exported as well as imported. For example, a heightmap generated as part of a terrain asset, might be exported and fine-tuned in an image editing program, then re-imported into the terrain generation tool, where the modified heightmap will affect existing 3d geometry. Similarly, if a dungeon is updated and re-exported, we should consider what workers further down the content pipeline will need to update game logic and re-apply art. By portraying our tools in a real content pipeline, we can focus the scope of our work. 

It is the authors' hope that this paper can contribute to a discussion about how we can improve MI tools for PCG and reach across the divide to people with a similar interest in the games industry.
%
\begin{acks}
  This work was funded by the EPSRC Centre for Doctoral Training in Intelligent Games \& Game Intelligence (IGGI) EP/L015846/1.
\end{acks}

%
\bibliographystyle{ACM-Reference-Format}
\bibliography{manifesto}

%

\end{document}